\documentclass[a4paper]{article}

\usepackage[english]{babel}    
\usepackage[utf8]{inputenc}

\usepackage{fancybox}
\usepackage{pdfpages}
\usepackage{stmaryrd}
\usepackage{amssymb}
\usepackage{amsmath}
\usepackage{amsthm}
\newtheorem{definition}{Definition}
\usepackage{mathabx}
\usepackage{amsfonts} 
\usepackage{stmaryrd}                   

\usepackage{todonotes}         
\usepackage{tikz}              
\usetikzlibrary{shapes,arrows} 

\usepackage{alltt}
\usepackage{url}
\usepackage{listings}
\newcommand{\jpdinl}[1]{\lstinline{#1}}

\usepackage{enumitem}          

\begin{document}

\newcommand\after[1]{\ensuremath{\circ #1}}

\newcommand{\forceindent}{\leavevmode{\parindent=1em\indent}}

\def\Name{\ensuremath{\text{\bf Name}}}
\def\OR{\ensuremath{\ |\ }}
\def\TO{\ensuremath{\rightarrow}}
\def\FROM{\ensuremath{\leftarrow}}
\def\LB{\ensuremath{\llbracket}}
\def\RB{\ensuremath{\rrbracket}}
\def\id{\ensuremath{\mathrm{id}}}

\newcommand\LIT[1]{\ensuremath{\text{\tt #1}}}
\newcommand\SLIT[1]{\ \LIT{#1}\ }
\newcommand\IF[3]{\LIT{if}\ #1\ \LIT{then}\ #2\ \LIT{else}\ #3}
\newcommand\APP[2]{#1\ #2}
\newcommand\CASE[4]{\LIT{case}\ #1:#2\ \LIT{of}\ #3 \TO #4}
\newcommand\INBR[1]{\ensuremath{\llbracket #1 \rrbracket}}
\newcommand\MGU[2]{\LIT{unify}(#1, #2)}
\newcommand\INVERT[1]{(\LIT{invert}\ #1)}

\newcommand\EvalWith[4]{\ensuremath{ \Delta#1 #2 \vdash #3 \downarrow #4}}
\newcommand\Eval[2]{\EvalWith{}{\Gamma}{#1}{#2}}
\newcommand\LaveWith[4]{\ensuremath{ \Delta#1 #2 \vdash #3 \uparrow #4}}
\newcommand\Lave[2]{\LaveWith{}{\Gamma}{#1}{#2}}

\newcommand\Because[4]{\ensuremath{\Delta{#1}\INBR{#2 \downarrow #3}\rightsquigarrow #4}}
\newcommand\Esuaceb[4]{\ensuremath{\Delta{#1}\INBR{#2 \uparrow #3}\leftsquigarrow #4}}

\newcommand\LinearHasType[4]{\ensuremath{\Delta#1 #2 \vdash #3 : #4}}
\newcommand\LinearEpytSah[4]{\ensuremath{\Delta#1 #2 \vDash #3 : #4}}
\newcommand\LinearBindTypes[4]{\ensuremath{\Delta#1 | #2 : #3 \Downarrow #4}}
\newcommand\LinearUnBindTypes[4]{\ensuremath{\Delta#1 | #2 : #3 \Uparrow #4}}

\newcommand\UPARROW[1]{$\uparrow$#1}
\newcommand\DOWNARROW[1]{$\downarrow$#1}
\newcommand\RIGHTARROW[1]{$\rightarrow$#1}
\newcommand\LEFTARROW[1]{$\leftarrow$#1}
\newcommand\TYPE[1]{$\tau$#1}
\newcommand\LinINFER[1]{$\Downarrow$#1}
\newcommand\LinUNFER[1]{$\Uparrow$#1}

\newcommand \BX [1]
  {\scriptsize\framebox{{\raisebox{0pt}[0.7\baselineskip][0.01\baselineskip]{\small #1}}}}

\newcommand\Axiom[2]
                 {\ensuremath{\text{\small #1}:\frac{\displaystyle}
                 {\displaystyle #2}}
                 }
\newcommand\InfOne[3]
                 {\ensuremath{\text{\small #2}:\frac{\displaystyle #1}
                 {\displaystyle #3}}
                 }
\newcommand\InfTwo[4]
                 {\ensuremath{\text{\small #3}:\frac{\displaystyle #1 \quad #2}
                 {\displaystyle #4}}
                 }
\newcommand\InfThree[5]
                 {\ensuremath{\text{\small #4}:
                     \frac{\displaystyle #1 \quad #2 \quad #3}
                          {\displaystyle #5}}
                 }
\newcommand\InfFour[6]
                 {\ensuremath{\text{\small #5}:
                     \frac{\displaystyle #1 \quad #2 \quad #3 \quad #4}
                          {\displaystyle #6}}
                 }
\newcommand\InfFive[7]
                 {\ensuremath{\text{\small #6}:
                     \frac{\displaystyle #1 \quad #2 \quad #3 \quad #4 \quad #5}
                          {\displaystyle #7}}
                 }


\title{Branching execution symmetry in Jeopardy by available implicit arguments
  analysis}

\author{Joachim Tilsted Kristensen$^1$, Robin Kaarsgaard$^2$\footnote{The second author is supported by DFF--International Postdoctoral Fellowship 0131-00025B.
}, \\and Michael Kirkedal Thomsen$^{1,3}$}

\date{$^1$ University of Oslo, Norway\\
$^2$ University of Edinburgh, UK\\
$^3$ University of Copenhagen, Denmark}

\maketitle


\begin{abstract}
  When the inverse of an algorithm is well-defined -- that is, when its
  output can be deterministically transformed into the input producing it --
  we say that the algorithm is invertible.  While one can describe an
  invertible algorithm using a general-purpose programming language, it is
  generally not possible to guarantee that its inverse is well-defined
  without additional argument. Reversible languages enforce deterministic
  inverse interpretation at the cost of expressibility, by restricting the
  building blocks from which an algorithm may be constructed.

  Jeopardy is a functional programming language designed for writing
  invertible algorithms \emph{without} the syntactic restrictions of
  reversible programming. In particular, Jeopardy allows the limited use of
  locally non-invertible operations, provided that they are used in a way
  that can be statically determined to be globally invertible. However,
  guaranteeing invertibility in Jeopardy is not obvious.

  One of the central problems in guaranteeing invertibility is that of
  deciding whether a program is symmetric in the face of branching control
  flow. In this paper, we show how Jeopardy can solve this problem, using a
  program analysis called available implicit arguments analysis, to
  approximate branching symmetries.

\end{abstract}

\section{Introduction}
\label{sec:introduction}


The interest in programs that can recover the inputs from a computed output
have long existed: from McCarthy's generate-and-test method~\cite{McCarthy:1956}
to the numerous inversion techniques associated with the \emph{reversible model
of computation}~\cite{Landauer:1961,Bennett:1973,Huffman:1959}.
Many languages have been designed to guarantee that programs describe
reversible algorithms, often by restricting \emph{which} programs are
allowed.  A notable such language is
Janus~\cite{LutzDerby:1986,YokoyamaGlueck:2007:Janus}, a reversible
imperative language which guarantees partial reversibility\footnote{Often
  (as in Janus) local invertibility only guarantees invertibility of partial
  functions. This comes from the fact that control structures (like
  conditional and loops) require assertion of specific values, and because
  procedures may fail to terminate.} by restricting programs to be sequences
of locally invertible statements.  Furthermore,
Theseus~\cite{JamesSabry:2014:RC} restricts programs to be a composition of
locally invertible surjective functions. Finally,
RFun~\cite{YokoyamaAxelsenGlueck:2012:LNCS,ThomsenAxelsen:2016:IFL} imposes
constraints that enforce function invertibility by enforcing a bidirectional
first-match policy on choice points (at runtime), and requiring programs to
be linear in their arguments. Doing so, is sufficient to guarantee
invertible algorithms while not restricting computational power beyond
R-Turing completeness\footnote{R-Turing completeness is Turing completeness
  restricted to programming languages (and hence Turing machines) defining
  only reversible programs.}. Both of these constraints can clearly be
checked statically: for the former, we may require the programmer to unroll
their program until input and output patterns are syntactically orthogonal,
and the latter can be enforced by linear
typing~\cite{Girard:1987,Wadler:1990} as has been shown for
CoreFun~\cite{JacobsenEtal:2018}, a simple typed version of RFun.

However, writing algorithms in a way that makes their reversibility evident
can be difficult, as it corresponds, in a certain way, to asking the programmer
to prove this property as they are writing the program. Writing programs in
these reversible languages requires some experience, and can in some cases be
notoriously hard.
The first attempts to this approach, was McCarthy's generate-and-test
algorithm~\cite{McCarthy:1956}. As this method is often infeasible in
practice, later research approached the problem using program
inversion~\cite{GlueckKawabe:2005:LR,HouEtal:2013:RC} or even
semi-inversion~\cite{Mogensen:2008,KirkebyGluck:2009}. Since these methods
all build on the conventional programming model, they may fail in cases
where a deterministic inverse does not exist.

The language Jeopardy~\cite{Kristensen:2022:IFLwip} has been designed with
inversion in mind. It can be seen as a combination of the above two approaches:
restricting the syntax enough to be able to give static inversion guarantees,
but relaxing the execution model enough to make programming as natural as
possible. It has a syntactic resemblance to your garden variety functional
programming language and exhibits the expected semantics for programs
running in the conventional direction.

However, not all algorithms describe bijective functions, and the problem of
deciding whether an algorithm is invertible is undecidable in general
following from Rice's Theorem. This means that the static analysis needed
to guarantee inversion of even simple Jeopardy programs is not
straightforward. In this work, we investigate the approach of approximating
global program invertibility by developing a data flow analysis that infers
the information necessary to make the approximation.

To be precise, in Section~\ref{sec:explain} we outline the problem by
providing an instructive program example. In Section~\ref{sec:specification}
we briefly outline the syntax and semantics of the Jeopardy programming
language; a more formal introduction to the language was presented at IFL
2022~\cite{Kristensen:2022:IFLwip}. In Section~\ref{sec:examples} we detail
the meaning of implicit arguments to functions that are inversely
interpreted.  In Section~\ref{sec:analysis} we provide an algorithm for
performing \emph{implicitly available arguments analysis} on Jeopardy
programs. Furthermore, in Section~\ref{sec:instructive-example} we run the
algorithm on the program example, and in Section~\ref{sec:discussion} we
discuss the implications of the result. Finally, in
Section~\ref{sec:conclusion} we conclude on the results.

A Haskell implementation of \emph{available implicit arguments analysis} for
Jeopardy can be found at:

{\scriptsize\url{https://github.com/jtkristensen/Jeopardy/blob/main/src/Analysis/ImplicitArguments.hs}}

\section{Branching symmetries and invertibility}
\label{sec:explain}

The extensional behavior of a program can be reasonably thought of as a
function mapping inputs to outputs~\cite{DanielssonNils2006Falr}. In this
perspective, the existence of an inverse program is analogous to the
existence of an inverse function. That is, a program $f : A \rightarrow B$
is invertible when a deterministic inverse program $f^{-1} : B \rightarrow
A$ exists, such that the functions they describe satisfy $f^{-1}\after{f} =
\id_{A}$ and $f\after{f^{-1}} = \id_{B}$.

At the extensional abstraction of mathematical functions, we cannot infer
any more about a function's behavior than that which may be derived from the
premises given by the function's provider. However, when we are presented
with a program, we can perform program analysis that inspect it
to gain insight into its properties.

One such analysis, called \emph{available expressions
  analysis}~\cite{NielsonFlemming2015PoPA}, produces a set of expressions
per program point that has already been computed when the point is reached
at runtime. One purpose to perform this particular analysis can be to
transform programs into equivalent programs that do not recompute
expressions when it is not necessary.

In Jeopardy we wish to decide the set of available expressions that could
have been implicitly provided as function arguments at particular call sites
in a program.  Providing extra arguments to functions do not necessarily
make those programs run more efficiently, but it allows us to infer more
precise things about branching inside function calls ahead of runtime. For
instance, consider the program given in Figure~\ref{fig:fib-program}.

\begin{figure}
\begin{lstlisting}{haskell}
data natural_number = [zero] [successor natural_number].

sum (m, n) =
  case m of
  ; [zero]        -> n
  ; [successor k] -> sum (k, [successor n]).

fibber (m, n) = (sum (m, n), m).

fibonacci_pair n =
  case n of
  ; [zero]        -> ([successor [zero]], [successor [zero]])
  ; [successor k] -> fibber (fibonacci_pair k).

fibonacci n =
  case fibbonacci_pair n of
  ; (_, nth_fibonacci_number) -> (nth_fibonacci_number, n).

main fibonacci.
\end{lstlisting}
\caption{Example program, computing fibonacci numbers.}
\label{fig:fib-program}
\end{figure}

\noindent
The main function of the program \jpdinl{fibonacci} takes a natural number
$n$ as its argument, and produces a pair containing the $n$'th Fibonacci
number together with $n$ itself. It does so by projecting from the function
\jpdinl{fibonacci_pair} that produces a pair containing the $n$'th Fibonacci
number together with its successor. The pair is computed by recursively
applying the \jpdinl{fibber} function, which transforms a pair of Fibonacci
numbers into the next pair. As dictated by the definition of the Fibonacci
sequence, \jpdinl{fibber} finds the next number in the sequence by summing
the two previous numbers.

Starting from the top, the function \jpdinl{sum}, which computes sums of
pairs of natural numbers, is \emph{not} invertible. To be precise, the
solution $4$ does not have a unique corresponding problem. In particular, it
can be the sum of $1$ and $3$, or it can be the sum of $0$ and $4$. In the
latter case, \jpdinl{sum} would simply return the $4$ by taking the first
branch of the case statement, and in the former case, it would take the
second branch and call itself recursively.

However, the output of \jpdinl{sum} is still uniquely determined by the
input, and so, inferring which branch was taken in each call to
\jpdinl{sum}, is sufficient for uniquely determining its input. For
instance, in the function \jpdinl{fibber} we never call \jpdinl{sum} without
also returning the first of its arguments as well. Thus, in the inverse
interpretation of \jpdinl{sum}, if the first argument is $0$, we are
done. Otherwise, the first argument is the successor of some natural smaller
number \jpdinl{k}, and we can inverse interpret the \jpdinl{sum} function
recursively.

Similarly, it is not trivial that \jpdinl{fibonacci_pair} is an invertible
algorithm, since you need to know that the second branch of the
case-statement will always be syntactically orthogonal to a pair of
ones. However, as the argument for \jpdinl{fibbonacci_pair} is directly
available in the output of \jpdinl{fibonacci}, it is clearly invertible in
the context of inverse interpreting the \jpdinl{fibonacci} function.

To summarize, obtaining the inverse of the entire \jpdinl{fibbonacci}
program corresponds to writing a program that takes as argument a pair
containing the $n$'th Fibonacci number together with $n$ itself and merely
gives back the $n$. The right projection is insufficient in terms of
correctly determining the problem for this solution, because its
interpretation in the conventional direction does not compose to an
identity. However, recovering information about branching \emph{is}
sufficient for inverse interpretation. Reversible programming languages that
enforce local invertibility, such as RFun and
CoreFun~\cite{YokoyamaAxelsenGlueck:2012:LNCS,JacobsenEtal:2018} simply
throw an error at runtime if branching does not comply with a
\emph{symmetric first match policy} for pattern matching, and
Theseus~\cite{JamesSabry:2014:RC} requires the programmer to account for
branching structure syntactically.

However, as we have seen in the \jpdinl{fibonacci} program example,
realistic programs often exhibit inter-procedural information that allows us
to recover the branching structure. It remains to show how this information
may be obtained systematically. In the remainder of this article, we concern
ourselves with doing just that.

\section{The Jeopardy Programming Language}
\label{sec:specification}

Jeopardy is a carefully designed first order functional language aimed at
expressing invertible algorithms while enabling concise program analysis and
dissemination. The main features are user-definable algebraic data types and
explicit function level program inversion. The full grammar can be found in
Figure~\ref{fig:syntax}.

\begin{figure}
\begin{center}
\begin{align*}
x          &\in \Name                                        &\text{(Well-formed variable names).}\\
c          &\in \Name                                        &\text{(Well-formed constructor names).}\\
\tau       &\in \Name                                        &\text{(Well-formed datatype names).}\\
f          &\in \Name                                        &\text{(Well-formed function names).}\\
p          &::= [c\ p_i] \OR x                               &\text{(Patterns).}\\
v          &::= [c\ v_i]                                     &\text{(Values).}\\
\Delta     &::= f\ (p : \tau_p) : \tau_t\ =\ t\ .\ \Delta    &\text{(Function definition).}\\
           &\OR \LIT{data}\ \tau\ =\ [c\ \tau_i]_j\ .\ \Delta &\text{(Data type definition).}\\
           &\OR \LIT{main}\ g\ .                             &\text{(Main function declaration).}\\
g          &::= f \OR \INVERT{g}                             &\text{(Inversion).}\\
t          &::= p                                            &\text{(Patterns in terms).}\\
           &\OR g\ p                                         &\text{(First order function application).}\\
           &\OR \CASE{t}{\tau}{p_i}{t_i}                     &\text{(Case statement).}
\end{align*}
\end{center}
\caption{The syntax of Jeopardy.}
\label{fig:syntax}
\end{figure}

Running a program corresponds to calling the declared \jpdinl{main} function
on a value provided by the caller in the empty context. Similarly, running a
program backwards corresponds to calling the main function's inverse on a
value provided by the caller in the empty context as well. Since an
application is a term, reasoning about inversion of terms \emph{is the same
as} reasoning about inversion of programs.

\begin{figure}
\begin{align*}
  \LB [c\ t_i] \RB_{\Delta[\LIT{data}\ \tau = [c \tau_i]_j]} &:= \CASE{t_i}{\tau_i}{p_i}{[c\ p_i]}\\
  \LB (t_1, t_2) \RB_\Delta &:= \LB [\LIT{pair}\  t_1\ t_2]\RB_\Delta\\
  \LB t_1 : t_2 \RB_\Delta &:= \LB [\LIT{cons}\ t_1\ t_2]\RB_\Delta\\
  \LB \LIT{[]} \RB_\Delta &:= [\LIT{nil}]\\
  \LB \APP{f}{t}\RB_{\Delta[f (\cdot : \tau) : \cdot = \cdot]} &:= \CASE{t}{\tau}{p}{\APP{f}{p}}\\
  \LB \LIT{let}\ p : \tau =\ t\ \LIT{in}\ t' \RB_{\Delta} &:=
  \CASE{t}{\tau}{p}{t'}\\
  \LB t' \RB_{\Delta[f (p_i : \tau_p) : \tau_t = t_i.]} &:= \LB t' \RB_{\Delta[f (x : \tau_p) : \tau_t = \CASE{x}{\tau_p}{p_i}{t_i}{}]}
\end{align*}
\caption{Disambiguation of syntactic sugar.}
\label{fig:sugar}
\end{figure}

The syntax of terms has been designed with the goal of providing program
analysis at the cost of making programs harder to read and write. In the
interest of writing intuitive program examples, we have therefore equipped
Jeopardy with a set of derived syntactic connectives that we have shown in
Figure~\ref{fig:sugar}.  Additionally, we may choose to omit type
annotations whenever these are not necessary, and use literal syntax for
natural numbers (\LIT{0}, \LIT{1}, \dots) to mean their data representations
(\LIT{[zero]}, \LIT{[successor [zero]]}, \dots).

\section{Implicit Arguments}
\label{sec:examples}

Recall from Section~\ref{sec:explain} that \jpdinl{sum} was not injective,
and thus its inverse was not well-defined. A na\"ive solution to this
problem is to automatically injectivise the program using, e.g., Bennett's
method~\cite{Bennett:1973}: that is, we keep a computation history of our
program, copy its result, and uncompute the history to be left with the
input and a copy of the result.  However, this is a very inefficient way of
doing invertible computing and would constantly generate extra unwanted
data. In many cases we can do better by first determining which inputs are
needed for uniquely deciding branching information bidirectionally. For
example, if we want to make \jpdinl{sum} invertible it suffices to copy one
of its inputs to the outputs as follows:

\begin{lstlisting}[escapeinside=\%\%]{haskell}
sum_and_copy_first (m, n) =
  case m of
  ; [zero]        -> (m, n)
  ; [successor k] ->
    case sum_and_copy_first (k, [successor n]) of
    ; (k, k%+%suc_n) -> (m, k%+%suc_n).

sum_and_copy_second (m, n) =
  case m of
  ; [zero]        -> (n, n)
  ; [successor k] ->
    case sum_and_copy_second (k, [successor n]) of
    ; (k%+%suc_n, [successor n]) -> (n, k%+%suc_n).
\end{lstlisting}

\noindent
To convince the reader that the branching symmetry is recoverable from the
transformed program, in \jpdinl{sum_and_copy_first} we are matching on
\jpdinl{m}, and \jpdinl{m} is embedded directly in the output. To see that
branching is symmetric in \jpdinl{sum_and_copy_second}, we need to show that
\jpdinl{k+suc_n} and \jpdinl{n} are different in the last branch, which we
can by co-induction since the recursive call returns a pair of successors of
\jpdinl{n}, or some larger structure that contains \jpdinl{n} from previous
calls.  Regardless of our choice of injectivisation of \jpdinl{sum}, we need
to store information from the previous call in order to make branching
symmetry decidable for the next, effectively by supplying a function (in
this case the inverse to \jpdinl{sum}) with extra arguments.

By producing specialised functions that take extra arguments in this way, we
can transform programs that contain the original functions into equivalent
programs that call the specialised function whenever the extra arguments are
available.  For instance, we might rewrite \jpdinl{fibber} from
Figure~\ref{fig:fib-program} as follows:

\begin{lstlisting}[escapeinside=\%\%]{haskell}
fibber_specialized_for_sum_and_copy_first (m, n) =
  case sum_and_copy_first (m, n) of
  ; (m, m%+%n) -> (m%+%n, m).
\end{lstlisting}

\noindent
This transformation depends on knowing which specialised versions of a
particular function it can use, and in turn, it depends heavily on knowing
what terms are available, or can be made available in the program point at
which the call happens. Because Jeopardy is a pure functional language, this
is simply all terms that appear in all paths to the program point in a call
graph with the programs main function as its entry point, and we already
know how to construct such a graph \cite{NielsonFlemming2015PoPA}.

In fact, we can do a little better, since we only need to require that a
term appears on every path that allows us to apply specialisation; though not
necessarily the same specialisation in every path. In this way, the goal of
our algorithm is as follows: for every function application in a program,
for every distinct path to that application from the main function, compute
what terms are available to be provided as implicit (extra) arguments.

\section{The Algorithm}
\label{sec:analysis}

To avoid having to deal with names, our algorithm performs an initial
annotation of the input program, where each program point is assigned a
unique integer label, as will be demonstrated for
Figure~\ref{fig:fib-program} in
Section~\ref{sec:instructive-example}. Furthermore, to make things more
concrete, we specify what it means to be a call-configuration, as specified
in Definition~\ref{def:call}.

\begin{definition}
  \label{def:call}
  A call-configuration is a 4-tuple $(c, f, A, I)$ , containing the name
  ``$c$'' of the function in which the call occurred, the (possibly
  inverted) function ``$f$'' being called, a set ``$A$'' containing the
  labels of the arguments to the function, a set ``$I$'' of available
  implicit arguments from previous calls in which the program is running at
  the time of the call.
\end{definition}

Now, achieving the goal presented in Section~\ref{sec:examples}, is
equivalent to answering the question: for each call in a program, what are
the possible configurations of the call?

We answer this question by solving a set of equations. Each equation, have a
fixed program-of-interest $\Delta$. We give the name $\mathcal F$ to the set
of function names defined in $\Delta$; The superset of $\mathcal F$ that
includes inversions (function names occurring under the keyword
\jpdinl{invert}), we call $\mathcal I$.  Furthermore, we assign the name
$\mathcal L$ to the set of labels of $\Delta$, and finally, we give the name
$\mathcal C \subseteq (\mathcal F \times \mathcal I \times \mathcal L \times
\mathcal L)$ to the set of possible call-configurations.

To produce all the possible configurations, we declare a function that
computes the closure of the call-configurations that are reachable from two
initial configurations\footnote{forward and backward from main.}:

\begin{align*}
  \text{configuration} : \Delta \TO \mathcal P (\mathcal C)
\end{align*}

\noindent
Its corresponding definition can be found in
Figure~\ref{fig:configurations}, where $\Delta$ has been extended with a
special top level function $\top$ and two special labels ``input'' and
``output'' for the arguments that should be provided by the entity that runs
the program. The equation, and the computations it depends, on are all
defined in this section.

\begin{figure}[ht]
  \begin{align*}
    \text{configurations}(\Delta[main\ g.]) &=
    \{ (\top, g, \{\text{input}\}, \emptyset)
    ,  (\top, \INVERT{g}, \{\text{output}\}, \emptyset)
    \}\\ & \cup \left( \bigcup_{c \in \text{configurations}(\Delta)}\text{call}(c)_{\Delta} \right)\\
  \end{align*}
\caption{The reachable call configurations from main.}
\label{fig:configurations}
\end{figure}

\noindent
In the last part of the definition of \text{configurations}, a function:
\begin{align*}
  \text{call} : \mathcal C \TO \mathcal P (\mathcal C)
\end{align*}
takes as argument, a configuration and returns all the possible
configurations that are reachable by calling the (possibly inverted)
function $f$ from that configuration, as defined in figure \ref{fig:calls}

\begin{figure}[ht]
\begin{align*}
  \mathrm{call}((c, f, A, I))_{\Delta[f p = t.]} &=
  \begin{cases}
    \text{term}\downarrow(f, (I \cup A) \setminus (\text{labels}(p,t)), t) &
    :\ \text{dir}(f) = \downarrow \\ \pi_1(\text{term}\uparrow(f,(I
    \cup A) \setminus (\text{labels}(p,t)), t)_\Delta) &
    :\ \text{otherwise}\\
  \end{cases}\\
\end{align*}
\caption{The reachable configurations from a given configuration.}
\label{fig:calls}
\end{figure}

Depending on the direction of execution, the reachable definitions are
defined by a inspecting the terms that constitute the terms and patterns
that define the body and argument of a function. We give the name $\mathcal
T$ for such terms, and further declare two functions:
\begin{align*}
  & \text{term}\downarrow : (\mathcal F,\mathcal I,\mathcal T) \TO \mathcal
  P (\mathcal C)\\ \text{and}\quad &\text{term}\uparrow : (\mathcal
  F,\mathcal I,\mathcal T) \TO \mathcal P (\mathcal C \times \mathcal L)
\end{align*}

\noindent
that compute the reachable configurations depending on said direction
$dir(f)$ in which the call is to be interpreted. We define these functions
in Figure~\ref{fig:terms}. They both return the set of configurations
reachable from their argument term $t$. However, the function interpreting
calls against the conventional direction returns the labels of available
expressions from ``the future'' as a means of definitional convenience.

In the case for patterns, both functions yield an empty set of
call-configurations since a pattern cannot contain an application. That is,
terms in patterns are syntactic sugar that we disambiguated in
Figure~\ref{fig:sugar}. The cases for function application yield the
configurations reachable as defined by the function ``call''. Regarding
case-statements, both functions return the collection of configurations
reachable in each branch.

\begin{figure}[ht]
\begin{align*}
  \text{term}\downarrow(f, I, \LB{} p \RB) &= \emptyset\\
  \text{term}\downarrow(f, I, \LB{} \APP{g}{p}\RB) &= \{(f, g,
  \text{labels}(p), I, \text{direction}(g))\}\\
  \text{term}\downarrow(f, I, \LB{} \LIT{case}\ t\ \LIT{of}\ p_j \TO t_j \RB_{j \in J}) &=
  \text{term}\downarrow(f, I, t)\\
  &\cup ( \bigcup_{j\in J} (\text{term}\downarrow((f, I \cup (\text{labels}(t) \cup
  \text{labels}(p_j)), t_i)))) \\
  \text{term}\uparrow(f, I, \LB{} p \RB)_\Delta &= \{(\emptyset, I \cup \text{labels}(p))\}
  \\
  \text{term}\uparrow(f, I, \LB{} \APP{g}{p}\RB^{l_1})_{\Delta[g q =
      t^{l_0}]} &= \{((f, g, \{l_0\}, I,
  \text{op}(\text{dir}(g))), \{l_1\}) \| c \in \mathrm{call}((f,
  g, \}\\
  \text{term}\uparrow(f, I, \LB{} \LIT{case}\ t\ \LIT{of}\ p_j \TO
  t_j \RB^l_{j \in J}) &=\\ \bigcup_{j \in J} \{(L_h , \{l\} L_t) \| & (L_s,
  L_j) \in \text{term}\uparrow(f, I, t_j) \\ & (L_h, L_t) \in
  \text{term}\uparrow(f, L_j \cup \text{labels}(p_j) , t)
  \}
\end{align*}
\caption{Call configurations, reachable in terms for each direction of
  interpretation.}
\label{fig:terms}
\end{figure}

The direction $dir$, and the opposite direction $op$, of a function call are
defined by their corresponding functions in Figure~\ref{fig:directions}, and
with that we are done defining the algorithm.
\begin{figure}[ht]
\begin{align*}
  \text{dir}(g) &=
  \begin{cases}
    \text{op}(\text{dir}(f)) &:\ g = \INVERT{f}\\
    \downarrow &: \text{otherwise}
  \end{cases}\\
  \text{op}(\downarrow) &=~ \uparrow\\
  \text{op}(\uparrow)   &=~ \downarrow
\end{align*}
\caption{Definition, the direction of a function call and its opposite
  direction.}
\label{fig:directions}
\end{figure}

Computing the set of possible reachable call configurations in a program
$\Delta$ now, corresponds to calling the function \jpdinl{configurations} on
$\Delta$. It does so by finding the least fixed point of the function
\jpdinl{call} from two initial configurations.  And, we know that this least
fixed point always exists (and thus the algorithm is well defined), from
observing that configurations can be given the structure of a complete
lattice by $(c,f,A,I) \sqsubseteq (c',f',A',I')$ iff $c=c'$, $f=f'$, $A=A'$,
and $I \subseteq I'$, with joins and meets of configurations (with the same
name, function, and label set) given by unions and intersections of
available implicit arguments. Further, it can be shown that the functions in
Figures~\ref{fig:configurations}, \ref{fig:calls} and \ref{fig:terms} are all
monotone with respect to this order, so it follows by Tarski's fixed point
theorem that the least fixed point we are looking for always
exists. Furthermore, since a program contains only finitely many labels, it
follows additionally that the analysis always terminates.

\section{Instructive Example}
\label{sec:instructive-example}

On a less theoretical note, let us look at an example, namely that of finding
the available implicit arguments at all call sites in the Jeopardy example from
Figure \ref{fig:fib-program}. This is the same as finding a minimal fixed
point for the equation for ``configurations($\Delta$)'', where

{\footnotesize
\begin{lstlisting}[escapeinside=\%\%, mathescape=true, language=Haskell]
%$\Delta=$%
  (data natural_number = [zero] [successor natural_number].

   sum (m%$^{1}$%, n%$^{2}$%)%$^{0}$% =
     (case m%$^{4}$% of
      ; [zero]%$^{5}$%         -> n%$^{6}$%
      ; [successor k%$^{8}$%]%$^{7}$% -> (sum (k%$^{10}$%, [successor n%$^{12}$%]%$^{11}$%))%$^{9}$%)%$^{3}$%.

   fibber (m%$^{14}$%, n%$^{15}$%)%$^{13}$% = ((sum (m%$^{19}$%, n%$^{20}$%)%$^{18}$%)%$^{17}$%, m%$^{21}$%)%$^{16}$%.

   fibonacci_pair n%$^{22}$% =
     (case n%$^{24}$% of
      ; [zero]%$^{25}$%          -> ([successor [zero]%$^{28}$%]%$^{27}$%, [successor [zero]%$^{30}$%]%$^{29}$%)%$^{26}$%
      ; [successor k%$^{32}$%]%$^{31}$% -> (fibber (fibonacci_pair k%$^{35}$%)%$^{34}$%)%$^{33}$%)%$^{23}$%.

   fibonacci n%$^{36}$% =
     (case (fibbonacci_pair n%$^{39}$%)%$^{38}$% of
      ; (_%$^{41}$%, nth_fibonacci_number%$^{42}$%)%$^{40}$% -> (nth_fibonacci_number%$^{44}$%, n%$^{45}$%)%$^{43}$%)%$^{37}$%.

   main fibonacci%$^{46}$%.)
\end{lstlisting}
}

\noindent
As the main function is \LIT{fibbonaci} we immediately get:
\begin{align*}
\mathrm{configurations}&(\Delta[main\ \LIT{fibbonaci}.]) \\=&~
  \{ (\top, \LIT{fibbonaci}, \{\text{input}\}, \emptyset)
  ,  (\top, \INVERT{\LIT{fibbonaci}}, \{\text{output}\}, \emptyset)
  \}\\ \cup&~ \left( \bigcup_{c \in \text{configurations}(\Delta)}\hspace{-10mm}\mathrm{call}(c)_{\Delta} \right)\\
\end{align*}

\noindent
And, if we focus on the last term, and unfold it one step, we get:
\begin{align*}
  \left(\bigcup_{c \in \text{configurations}(\Delta)}\hspace{-10mm}\text{call}(c)_{\Delta} \right) = \mathrm{call}((\top, \LIT{fibbonaci}, \{\text{input}\}, \emptyset))_\Delta \cup \left(\bigcup_{c \in \text{configurations}(\Delta)}\hspace{-10mm}\text{call}(c)_{\Delta} \right)
\end{align*}

\noindent
If we again restrict our attention to the ``call'' part, the result is:
\begin{align*}
  \mathrm{call}((\top, &\LIT{fibbonaci}, \{\text{input}\}, \emptyset))_{\Delta[\LIT{fibbonaci n}^{36} = t]}\\ =~
  &\mathrm{term}\downarrow(\LIT{fibbonaci},\{\text{input}\} \setminus \{36\}, t)\\ =~
  &\mathrm{term}\downarrow(\LIT{fibbonaci},\{\text{input}\}, \LB{} \LIT{case}\ t^{38}\ \LIT{of}\ p_i^{40}\ \rightarrow\ t^{43}_i \RB{})\\=~
  &\mathrm{term}\downarrow(\LIT{fibbonacci}, \{\text{input}\}, t^{38} )\\ \cup~
  &\mathrm{term}\downarrow(\LIT{fibbonacci}, \{\text{input}\} \cup \mathrm{labels}(t^{38}) \cup \mathrm{labels}(p_i^{40}), t_i^{43})\\ =~
  & \{(\LIT{fibbonacci}, \LIT{fibbonacci\_pair}, \{39\}, \{\text{input}\})\} \cup \emptyset
\end{align*}

\noindent
Finally, we get
{\allowdisplaybreaks
\begin{align*}
\mathrm{configurations}&(\Delta[main\ \LIT{fibbonaci}.]) \\
  =
  \{ & (\top, \LIT{fibbonaci}, \{\text{input}\}, \emptyset), \\
     & (\top, \INVERT{\LIT{fibbonaci}}, \{\text{output}\}, \emptyset), \\
     & (\LIT{fibbonacci}, \LIT{fibbonacci\_pair}, \{39\}, \{\text{input}\})
  \}\\
  \cup & \left( \bigcup_{c \in \text{configurations}(\Delta)}\hspace{-10mm}\mathrm{call}(c)_{\Delta} \right)\\
  =
 \{ & (\top, \LIT{fibbonaci}, \{\text{input}\}, \emptyset),\\
    & (\top, \INVERT{\LIT{fibbonaci}}, \{\text{output}\}, \emptyset),\\
    & (\LIT{fibbonacci}, \LIT{fibbonacci\_pair}, \{39\}, \{\text{input}\}),\\
    & (\LIT{fibbonacci\_pair}, \LIT{fibber\_pair}, \{35\}, \{\text{input}, 39, 22, 23, 24, 31, 32\}),\\
    & (\LIT{fibbonacci\_pair}, \LIT{fibber}, \{34\}, \{\text{input}, 39, 22, 23, 24, 31, 32 \})
 \}\\
 \cup & \left( \bigcup_{c \in \text{configurations}(\Delta)}\hspace{-10mm}\mathrm{call}(c)_{\Delta} \right)\\
  =
 \{ & (\top, \LIT{fibbonaci}, \{\text{input}\}, \emptyset),\\
    & (\top, \INVERT{\LIT{fibbonaci}}, \{\text{output}\}, \emptyset),\\
    & (\LIT{fibbonacci}, \LIT{fibbonacci\_pair}, \{39\}, \{\text{input}\}),\\
    & (\LIT{fibbonacci\_pair}, \LIT{fibber\_pair}, \{35\}, \{\text{input}, 39, 22, 23, 24, 31, 32\}),\\
    & (\LIT{fibbonacci\_pair}, \LIT{fibber}, \{34\}, \{\text{input}, 39, 22, 23, 24, 31, 32\}),\\
    & (\LIT{fibber}, \LIT{sum}, \{18, 19, 20\}, \{\text{input}, 39, 22, 23, 24, 31, 32, 13, 14, 15, 21\}),\\
    & (\LIT{sum}, \LIT{sum}, \{10, 11, 12\}, \{0, 1, 2, 4, 7, 8\})
  \}\\
  \cup & \left( \bigcup_{c \in \text{configurations}(\Delta)}\hspace{-10mm}\mathrm{call}(c)_{\Delta} \right)\\
\end{align*}}
Here the additional call-configurations in the last part are derived in a
similar fashion to that which we did when we focussed on the equation for
``call'' in the conventional direction, and in the interest of saving space
in the paper, deriving the call configurations for
\INVERT{\LIT{fibbonacci}} is left as an exercise for the reader.

\section{Discussion}
\label{sec:discussion}

The transformation suggested in Section~\ref{sec:explain} requires us be
able to infer a particular set of call-configurations. We have designed and
implemented an algorithm for inferring said configurations. The algorithm
works well for finding implicitly available arguments in function calls, but
is limited in scope, to the configurations that do not reach beyond a single
step of recursion in its search for implicit arguments. However, a single
step is not a theoretical limit. It is easy to imagine a generalized
algorithm that infers implicitly available arguments up to a fixed depth,
but less so for arbitrary depth recursion.

In Section \ref{sec:instructive-example}, we have seen how to compute the
implicitly available arguments at each program point in an example program
that computes the Fibonacci numbers. The result of the analysis allows us to
replace functions for which we cannot decide branching symmetry with
specialized variations for which we can. For instance, at program point
$17$, \jpdinl{fibber} calls \jpdinl{sum} in a context where its first
argument \jpdinl{m} is implicitly available in both directions (from program
point $21$) as witnessed, for the conventional direction, by the tuple
\begin{align*}
  (\jpdinl{fibber}, \jpdinl{sum}, \{\dots\}, \{\dots, 21,\dots\})
\end{align*}
And so, we can make \jpdinl{fibber} invertible by replacing the
non-invertible function call to \jpdinl{sum} with a call to the invertible
specialization \jpdinl{sum_and_copy_first} from Section \ref{sec:examples},
which in turn allows us to invert \jpdinl{fibber}, as demonstrated in the
example function \jpdinl{fibber_specialized_for_sum_and_copy_first} also in
Section \ref{sec:examples}.

The problem with regards to recursion, is that the set of implicitly
available expressions' labels $I$ in a configuration, according to
Definition \ref{def:call}, corresponds to the terms that were ``available
from previous calls''. So, in a circular call structure (like recursion) it
is not possible to see the difference in $I$ between a term bound in this
call, or the previous.  In the example program from
Section~\ref{sec:explain}, this is not a problem, because deciding if
e.g. \texttt{sum} is symmetric with respect to branching, only
relies on implicit arguments either from the previous call to
\texttt{fibber} or \texttt{sum} itself. However, one could imagine a
scenario, where a term actually exhibits branching symmetry even though our
analysis does not find sufficient information to say so.

\section{Conclusion}
\label{sec:conclusion}

We have designed a program analysis for statically inferring the expressions
that are available as implicit arguments in function calls in the Jeopardy
programming language. The current formulation of the algorithm can be
implemented in about 200 lines of fairly readable and maintainable Haskell
code. Our implementation makes use of the nifty Reader-Writer-State monad,
which turns out to be suitable for threading around the program $\Delta$, as
well as keeping track of termination conditions etc.

In the near future, we expect to develop a program transformation, that
rewrites Jeopardy programs where branching symmetry is not syntactically
apparent, into programs where it is. To give an intuition, we have provided
an example in Section~\ref{sec:examples}, of applying such a transformation
to the function \LIT{fibber} from the program example in
Section~\ref{sec:explain}.

The reasons for wanting to design and implement such a transformation is
twofold. As mentioned earlier, the main motivation is to enable static
analysis for deterministic backwards branching execution inference.  But
such a transformation also has similar motivation to that of compiling
programs rather than interpreting them. That is, a jeopardy interpreter
should not perform this analysis every time a function is called, or thread
around the implicit arguments at runtime, it should transform the program
into an equivalent program where implicit arguments have been explicitly
provided when necessary.

Static program analysis is almost always an
approximation, and cyclic call structures is where you will find the
limitations of the available implicit arguments analysis. However, it is not
hard to imagine a definition of call-configurations, that include (finitely
many) layers of cyclic references to previous calls. And, it is unclear at
the time of writing, if doing so will by useful in practice.

\bibliographystyle{abbrv}
\bibliography{sample-base}


\end{document}